# The magnetic and transport properties of edge passivated silicene nanoribbon by Mn atoms

Changpeng Chen [a, b,*],  Ziqing Zhu[b],  Dace Zha [b], Meilan Qi [b] ,Jinping Wu [a]
a. Material Science and Chemistry Engineering College, China University of Geosciences, Wuhan, 430074, PRChina
b.School of science, Wuhan University of Technology, Wuhan 430070 PRChina

Abstract：The effect of chemical doping on the ZSiNRs with Mn as passivating element replacing H atoms at one edge are investigated by first-principles calculations.The structures optimized in the typical ferromagnetic and antiferromagnetic coupling show that the system leads to an AFM state and achieve half-metallic properties.Also,our first principle approach based on the Keldysh non-equilibrium Green's function method gives the spin-dependent transport properties of the device. When the system changes from parallel to antiparallel configuration. the spin-up current increases rapidly while the spin-up current is still depressed. Further, it is found that the system is a quite good spin filtering device with nearly 80% spin filtering efficiency at a wide bias voltage region and therefore suitable for applications. The mechanisms for these phenomena are proposed in detail.

**1. Introduction**

   Owing to the remarkable structural and electronic properties and large potential applications in nanoelectronic devices, recently, silicene have attracted increasingly interest of many researchers.It seems that silicene would be a good replacement of graphene not only due to its graphene-like features but also its compatibility to existing silicon-based electronic devices [1]. Recent experimental work has demonstrated the synthesis of silicene sheets and silicon nanoribbons [2–5] grown on silver substrates. The experimental realization of silicene has opened a new avenue for exploring its properties and potential applications [6–12]. Hydrogen-passivated zigzag silicene nanoribbons (ZSiNRs) are currently being investigated intensely due to its potential applications in nanodevices e.g the field-effect transistors. And it is found that hydrogenation converts the intrinsic  bilayer silicene, a strongly indirect semiconductor, into a direct-gap semiconductor with a widely tunable band gap.[13] The electronic structure and transport properties of pristine zigzag SiNRs (ZSiNRs)[14–16] are similar to those of zigzag graphene nanoribbons (ZGNRs) [17],where all the edge silicon atoms on the same edge have the same magnetic moment. As is well known, there are several main approaches to change the properties of ZGNRs : modifying their edges [18]; doping with impurities in different sites [19–21];applying electric field [22, 23]and so on. Since Si materials can be doped by boron (B) or nitrogen (N) atoms. Luan H X. et al [24]studied the magnetic properties of silicene nanoribbons doped by the boron/nitrogen (B/N)bonded pair and found the introduction of the B/N pair results in a transition from nonmagnetic to spin-polarized states. Dong Y J et al [25] studied the spin-dependent electronic structures of aluminum doped zigzag silicene nanoribbons and found when ZSiNRs are substitutionally doped by a single Al atom on different sites in every three primitive cells,they become half-metallic. De Padova P[26] investigated the growth of Mn nanostructures on a one-dimensional grating of silicon nano-ribbons at atomic scale by means of scanning tunneling microscopy and found that the Mn atoms show a preferential adsorption site on silicon atoms,forming one-dimensional nanostructures.

Motivated by the route advances with the recent experimental and theoretical works of the SiNRs, in this article. we focus on the effect of chemical doping on the SiNRs with Mn as a passivating element replacing all the H atoms at one ege. the magnetic and transport properties are investigated in details. Our results show that the embedded Mn-passivations serve as a diluted magnetic semiconductor in the SiNRs. The spin current polarization is extracted for various bias and the transmission spectrum is depicted in detail, revealing quite considerable large spin-filtering effect. In this way we point out the applicability of Mn-passivated SiNRs in nanoscale spintronic devices.

## 2. Computational method and details

We perform the first principle calculations within the spin unrestricted density functional theory (DFT) based on the density functional method and ultrasoft pseudopotentials. All the data were performed with Atomistix Tool Kit-VirtualNanoLab(ATK-VNL) simulation package [27]. In this calculation,we considered an zigzag silicene nanoribbons (width $N = 6$ )with unit cells in a periodic box so as to make sure it is infinite in the one dimension.To ensure better accuracy,the calculations are performed with a plane wave cutoff energy of 150 Ry. To optimize the crystal structure and obtain the self-consistent electronic structure, brillouin zone integration was carried out at $5 \times 5 \times 5$ k-points. and $12 \times 12 \times 12$ k-points were used to obtain the density of states (DOS). The atomic positions were fully relaxed to change their position until the criteria of an energy convergence reached $2 \times 10^{-4}$ Ry. Particularly we model all the nanoribbons periodic along z axis. In order to prevent the artificial coulomb interaction between ribbon and its periodic images, we set a vacuum space of 10 Å in the non periodic (x and y) directions. Above all. we explicitly calculate the related value of the Mn-passivated zigzag silicene nanoribbons for the purpose of obtaining accurate results for the magnetic and Transport properties of the system.

## 3. Results and discussions

3. 1. The structural and magnetic properties

Firstly, we investigated the structure of the one-edge Mn-passivated silicene nanoribbons. The structure shown in Fig. 1 depicts that one edge of the H atoms are replaced by Mn atoms. The optimized structure after the passivation of the Mn atoms shows that the Si-Mn bond length at the edge is 2.13Å,undergoes a reduction of 0.18 Å after relaxation. The Si−Si bond length near the doped site increases to 2.31Å on the contrary. In the pure silicene nanoribbons, the overlapping of the $3p_z$ orbitals at the edge make the formation of π bonds, while in the Mn− passivation structure, these delocalized bonds are saturated by d orbitals of Mn and the Si-Si bonds at the edge altered to double bonds consequently. The angle of the Mn−Si−Si bond is 109.4°, close to the angle of free-standing silicene nanoribbons.

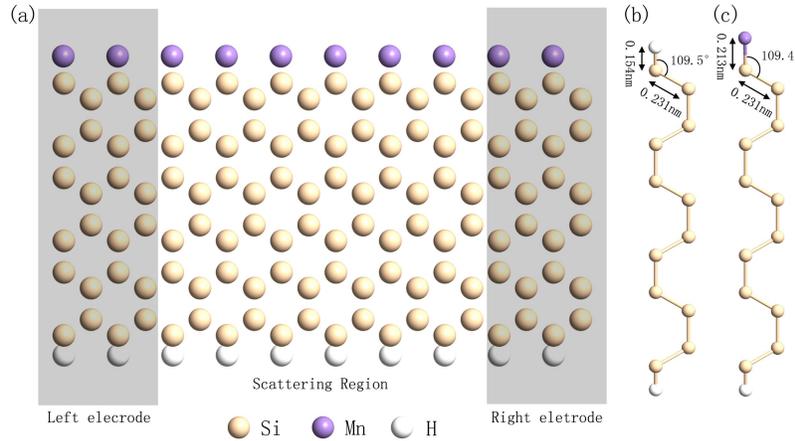

Fig. 1. （a）Structures of the Mn-passivated zSiNRs. the rectangle shows the optimized structural arrangement of the system in which one edge of the H atoms are entirely replaced by Mn atoms. (b) The fundamental crystal parameters on bond length and bond angle of the pure zSiNRs. (c) The fundamental crystal parameters on bond length and bond angle of the Mn-passivated zSiNRs. The yellow purple and white spheres represent Si. Mn and H atoms , respectively.

Next,we start our analysis by looking at the magnetic properties of the Mn-passivated zSiNRs. Fig. 2(a) sketches the spatial spin-density distribution of the structures. it shows that the magnetism is mainly localized around the Mn atoms. and there is no magnetism induced on the Si atoms. Also, references goes that the Mn doped armchair graphene nanoribbon have strong magnetic [28]. Thus it is possible for the unpaired 3d electrons in the unsaturated Mn sites to contribute to the expected magnetism. We consider two normal typical magnetic configurations in the zigzag armchair silcene ribbons: (i)Ferromagnetic(FM) coupling; (ii)Antiferromagnetic(AFM)coupling. For the purpose to find the stable magnetic state,we calculated the energies of FM and AFM states.The result show that the AFM states lies 0.56 meV lower in energy than FM states, which indicates that the introducing of the Mn-impurities on the ZSiNRs leads to an AFM state. Then we present the spin-polarized band structure of the Mn-passivated SiNRs at AFM coupling configurations in the Fig. 2(b) in order to understand and analyze their metallic or half-metallic behavior. As is well known, silicene is found to be a gapless semiconductor,the bonding π and antibonding π* bands crossing only at K points in the hexagonal brillouin zone [29]. As shown in Fig. 2(b), in this band structure. the spin-up bands cross the Fermi level while the spin-down bands have a mini gap of 0.16 eV,which indicates the half-metallic character of this configuration . In this case, the electronic states are non degenerate due to the broken symmetry of the ribbon. This is similar with the E. J. Kan's research which found that the boundary chemically modified can achieve half-metallic properties[30]. The broken degeneracy of electronic states constitutes the possibility of spin polarization and filtering the opposite spin electrons.

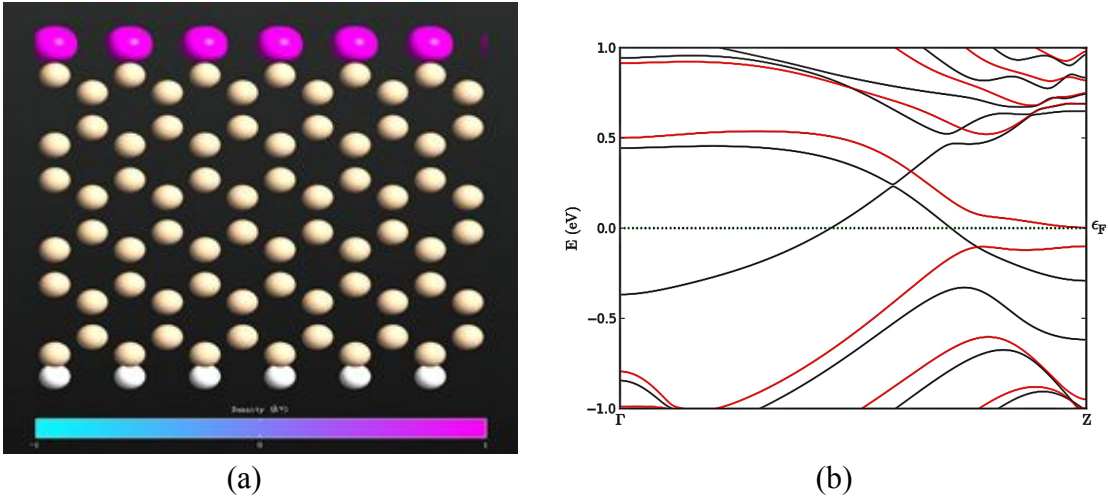

(a) (b)

Fig.2 (a) Spin density of the Mn-passivatied ZSiNRs. As the colors change from the left(blue) to the right(red), the spin density increase from zero to highest. (b) Band structures of model in AFM states. Spin up (down) states correspond to black (red) lines. The Fermi level is shown by dashed line at 0 eV. (c)Total density of states (DOS) for Mn-passivated SiNRs. Spin up (down) states correspond to black (red) lines.

3. 2. Transport properties

Firstly, we construct a two probe system of the left and right electrodes shown in Fig. 1(a)and consider two spin configurations: (i)P(parallel) configurations;(ii) AP(antiparallel) configuration. In P configuration both left and right electrodes are spin-up polarized while in AP configuration the two electrodes have inverse spin polarization directions. Using external magnetic field we can get these two configurations. The spin density of the distribution of the Mn-passivated SiNRs under zero bias for P and AP configurations is sketched in Fig. 4(a) and (b), respectively. A bias is applied between the two electrodes. From the figure one can see that the main magnetic distributions around the Mn impurities.The I–V characteristic for the Mn-passivated SiNRs are shown in the Fig. 5(a)and (b) where one can observe that the spin-up and spin-down currents are very close in P configuration, that is to say, the spin-filtering effect is particularly weak. Interestingly, while the magnetic configuration of the system is changed into AP configuration. the spin-up current increases rapidly after the bias is applied but the spin-down current is restrained significantly. Therefore. the current curved lines of the spin-up and spin-down states are obviously separated and perfect spin-filtering effect can be achieved. In order to assess the spin-filtering effect, the bias-dependent spin filter efficiency (BDSFE) for the AP configuration at zero bias and finite bias are calculated in Fig. 6. The BDSFE is defined in terms of the spin-resolved current using the formula as follows:

$$BDSFE = \frac{(I_{up} - I_{down})}{(I_{up} + I_{down})} \quad (1)$$

where $I_{up}$ and $I_{down}$ are the spin up current and the spin down current. respectively. From the curve graph. we noticed that the SFE value reaches nearly 80% at a bias voltage range from 0 to 0. 2 V for the AP configuration in the device. Especially, the high spin spin-filtering efficiency can be kept in a large bias region.

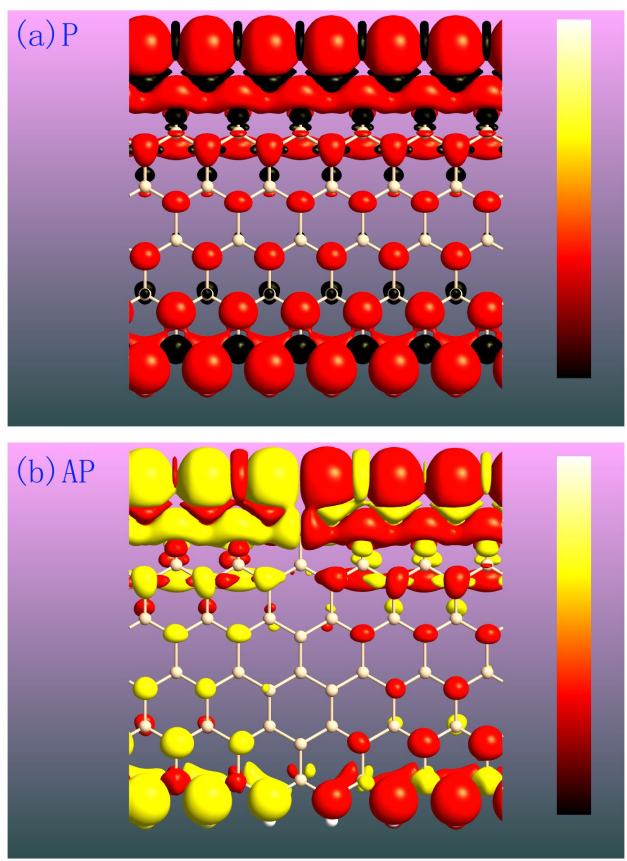

Fig. 4. (a), (b) Spin density of the model with P and AP spin configurations. respectively. As the colors change from the bottom(black) to the top(white), the spin density increase from zero to highest.

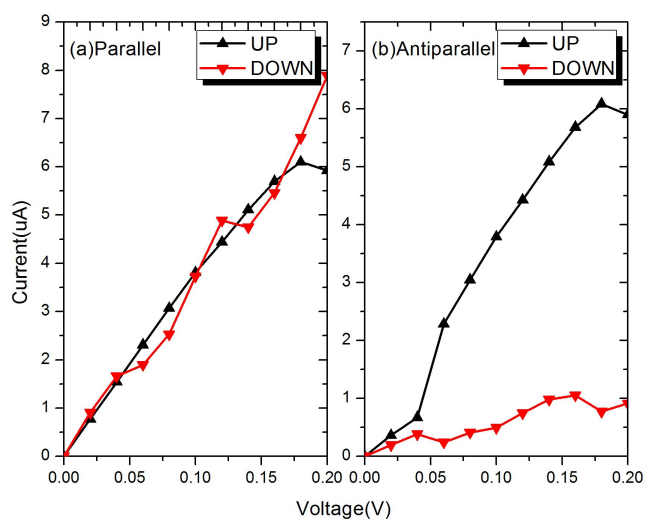

Fig. 5. (a), (b) Calculated current as a function of the applied bias for Mn-passivated SiNRs with P and AP spin configurations, respectively.

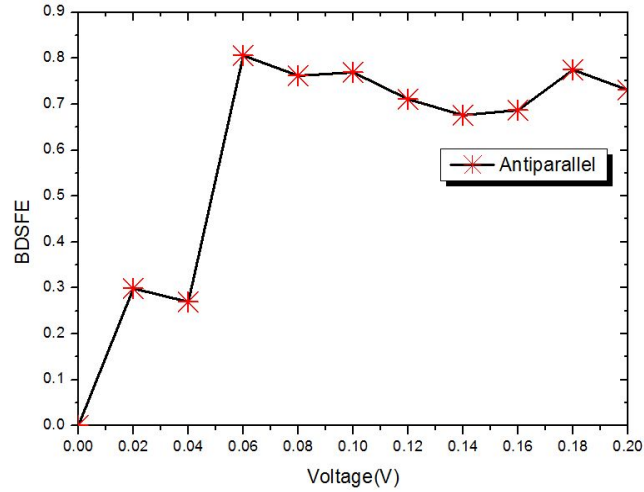

Fig. 6. the BDSFE of the system change with the applied bias for the AP configurations.

To further explain the origin of the spin-filtering effect in the system. Fig. 7(a) and (b) shows a series of spin-up and spin-down transmission spectra for the system with AP configurations at finite biases from 0 to 0. 2 V with a 0. 01 V interval. It's a definitely intuitive representation of electronic structures and quantum transport behaviors of a system, and it's generally accepted that the transmission spectrum should show distinctive feature in the spin up and spin down channels for an efficient spin filter. In the graph. the color shows the value of transmission coefficient. in which red indicates the largest value and black means almost zero. The two gradient white solid lines show bias window. In this figure. both the shapes of the transmission spectrum and the location of transmission peak change as the bias voltages vary. leading to different I–V characteristics up on the Landauer–Büttiker formula. To be specific, with the bias increasing. there is an obvious increase of transmission coefficients within windows. which can be excited to contribute to the currents. Furthermore. it is found that there are smaller spin-down transmission coefficients than spin-up transmission coefficients in bias.

The more detailed explanation on the spin-dependent transmission spectrum of the Mn-passivatied ZSiNRs is proposed in Fig. 8. And we take the AP configurations of the system at 0. 2 V s as an example. Furthermore. the corresponding molecular orbitals are also presented. As we know. the magnitude of transmission coefficients is related to the number of the molecular orbital and the delocalization degree of the molecular orbital. [31]. And the electronic transport in the scattering region depends on the frontier molecular orbitals to a great extent. Particularly. the HOMO(the highest unoccupied molecular orbital)and LUMO(the lowest unoccupied molecular orbital) which lie near the Fermi level are critical for the molecular devices. From Fig. 8(a). we can find that more molecular orbitals appear in the spin-up state compared with the spin-up state in bias. That is to say. there opens more tunneling transport channels in the spin-up state. leading the electrons injected from the electrode having a high possibility to flow through the central scattering region. Therefore. there are relatively high transmission peaks near the Fermi level. Oppositely. there are only LUMO in the bias window for the spin-down state and it must be insulating. The HOMO–LUMO gap noted from the data is much larger than that in the spin-up state. and conferences go that the lager the HOMO–LUMO gap. the more difficulty to transfer the electron [32][33], this mean the system in the spin-down state lead very low conductivity of electron transfer which is consistent with the conclusion from the transmission spectrum. Further, these results point that the perfect spin-filtering effect occurs.

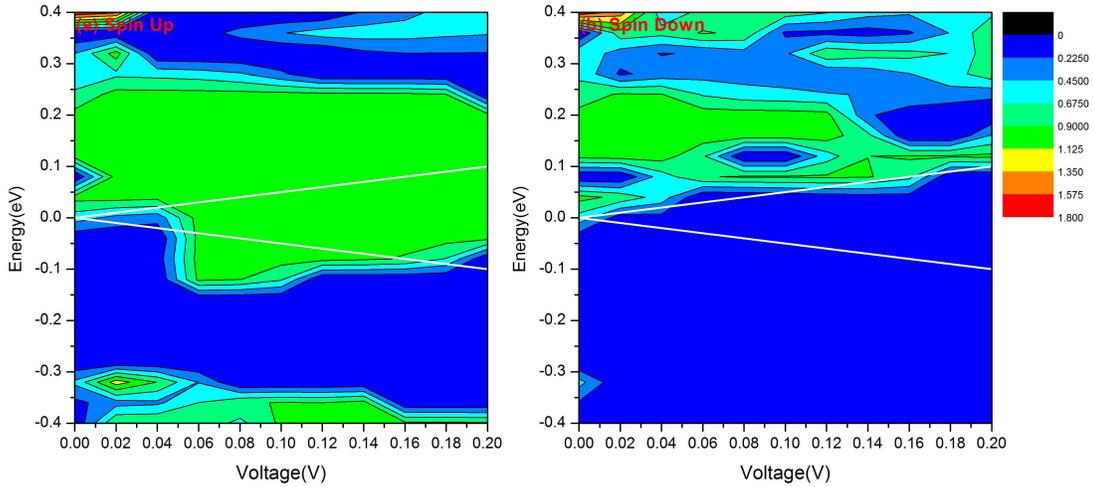

Fig. 7 (a) and (b) The total Transmission Spectrum in the AP configurations of the system at a series of biases from 0 to 0. 2 V with a 0. 01 V interval for spin up and spin down state. respectively. The Fermi level is set to zero.

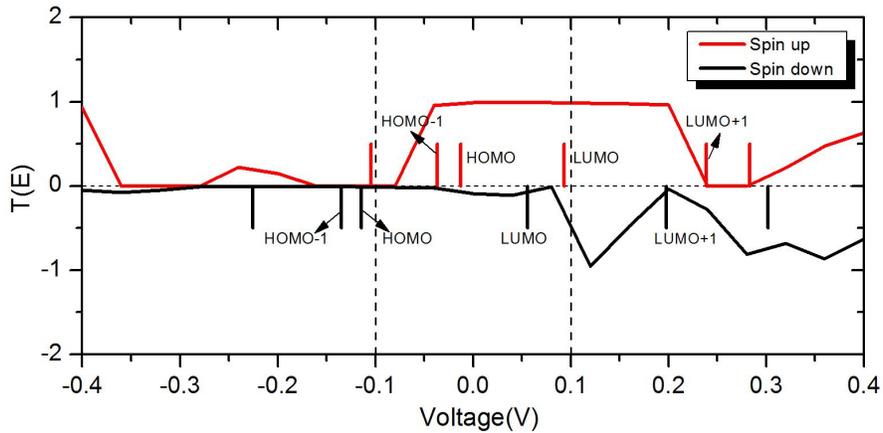

Fig. 8. Spin-dependent transmission spectrum of the system at 0. 06V. Spin up (down) states correspond to black (red) lines. The Fermi level is shown by dashed line at 0 eV.

## 4. Conclusion

In conclusion, applying the first-principles density functional calculations, we have investigated the magnetic and transport properties of the SiNRs with Mn as passivating element by replacing all the H atoms at one edge. We found that the Mn-passivated

ZSiNRs system is a good spin filtering device with nearly 80% spin filtering efficiency at a wide bias voltage region which is suitable for nano spintronics applications. Much tunneling transport channels have been opened in the spin-up state with quite high transmission coefficients while the channel gets suppressed in the spin-down state,leading very low conductivity of electron transfer. Therefore. our research work possesses the potential value for achieving the high-performance spin filter and spin valve.

**Acknowledgements**

The authors would like to acknowledge the support by the Project 61177076 supported by National Natural Science Foundation of China and NSAF (Grant No. U1330111).

**Reference**
[1] P. De Padova, O. Kubo, B. Olivieri, C. Quaresima, T. Nakayama, M. Aono, and G. Le Lay, Nano Lett. 12, 5500 (2012).
[2] C. Leandri, G. Le Lay, B. Aufray, C. Girardeaux, J. Avila, M.E. Dávila, M.C. Asensio, C. Ottaviani, A. Cricenti, Surf. Sci. 574, L9 (2005).
[3] B. Aufray, A. Vizzini, S. Oughaddou, H. Leandri, C. Ealet, G. Le Lay, Appl. Phys. Lett. 96, 183102 (2010).
[4] M. Davila, A. Marele,P. De Padova, Nanotechnology 23, 385703 (2012).
[5] D. Topwal, A. Kara, C. Carbone, P. Moras, G. Le Lay, C. Ottaviani, Applied Physics Letters. 96, 261905 (2010).
[6] S. Cahangirov, M. Topsakal, E. Aktürk, H. Sahin, S. Ciraci, Phys. Rev. Lett. 102, 236804 (2009).
[7] H. Sahin, S. Cahangirov, M. Topsakal, E. Bekaroglu, E. Akturk, R. T. Senger, S. Ciraci, Phys. Rev. B. 80, 155453 (2009).
[8] C. C. Liu, W. X. Feng, Y. G. Yao, Phys. Rev. Lett. 107, 076802 (2011).
[9] C. C. Liu, H. Jiang, Y. G. Yao, Phys. Rev. B. 84, 195430 (2011).
[10] M. Ezawa, Phys. Rev. Lett. 109, 055502 (2012).
[11] J. Sivek, H. Sahin, B. Partoens, and F. M. Peeters, Phys. Rev. B. 87, 085444 (2013).
[12] P. De Padova, P. Perfetti, B. Olivieri, C. Quaresima, C. Ottaviani, G. Le Lay, Journal of Physics: Condensed Matter. 24, 223001 (2012).
[13] B. Huang, H. Deng, H. Lee, M.Yoon, B. Sumpter, F. Liu, S. Smith, S.Wei, Phys. Rev. X. 4, 021029 (2014).
[14] J. Kang, F. Wu and J. Li, Appl. Phys. Lett. 100, 233122 (2012).
[15] D. Zha, C. Chen, J. Wu, Solid State Communications. 219, 21 (2015).
[16] D. Zha, C. Chen, J. Wu, Int. J. Mod. Phys. B. 29, 1550061 (2015).
[17] Z. Li, H. Qian, J. Wu, B.-L. Gu and W. Duan, Phys. Rev. Lett. 100, 206802 (2008).
[18] Z. Li, J. Yang and J. G. Hou, J. Am. Chem. Soc. 130, 4224 (2008).
[19] Y. Li, Z. Zhou, P. Shen and Z. Chen, ACS Nano. 3, 1952 (2009).
[20] X. H. Zheng, X. L. Wang, T. A. Abtew and Z. Zeng, J. Phys. Chem. C. 114, 4190 (2010).
[21] B. Huang, H. J. Xiang, S. Wei, Phys. Rev. Lett. 111, 145502 (2013).
[22] Y.-W. Son, M. L. Cohen and S. G. Louie, Nature 444, 347 (2006).

[23] E.-J. Kan, Z. Li, J. Yang and J. G. Hou, Appl. Phys. Lett. 91, 243116 (2007).
[24] H. X. Luan, C. W. Zhang, F. B. Zheng, J. Phys. Chem. C. 117, 13620 (2013).
[25] Y. J. Dong, X. F. Wang, P. Vasilopoulos, Journal of Physics D: Applied Physics. 47, 105304 (2014).
[26] P. De Padova, C. Ottaviani, F. Ronci, Journal of Physics: Condensed Matter. 25, 014009 (2013).
[27] M. Brandbyge, J. L. Mozos, P. Ordejon, J. Taylor, K. Stokbro, Phys. Rev. B. 65, 165401 (2002).
[28] N. Gorjizadeh, Y. Kawazoe, Materials Transactions. 49, 2445 (2008).
[29] S. Cahangirov, M.Topsakal, E. Aktürk, H. Sahin, S. Ciraci, Phys. Rev. Lett. 102, 236804 (2009).
[30] E.J. Kan et al., J. Am. Chem. Soc. 130, 4224 (2008).
[31] G.P. Tang, J.C. Zhou, Z.H. Zhang, X.Q. Deng, Z.Q. Fan, Carbon 60, 94 (2013).
[32] J. M. Seminario, A. G. Zacarias, J. M. Tour, J. Am. Chem. Soc. 122, 3015 (2000).
[33] M.-Q. Long, K.-Q. Chen, L.L. Wang, B. S. Zou, Z. Shuai, Appl. Phys. Lett. 1, 233512 (2007).